\documentstyle[psfig]{lamuphys}
\makeatletter
\let\chapter\hid@chapter
\makeatother
\begin{document}
\pagenumbering{arabic}

\title{Hubble Deep Fever: A faint galaxy diagnosis}

\author{Simon P. Driver}
\institute{School of Physics, University of New South Wales
Sydney, NSW 2052, Australia} 

\maketitle

\begin{abstract}
The longstanding faint blue galaxy problem is gradually subsiding as a result
of technological advancement, most notably from high-resolution Hubble Space 
Telescope imaging. In particular two categorical facts have recently been 
established, these are:

~

\noindent
1) The excess faint blue galaxies are of irregular morphologies,

~

and,

~

\noindent
2) the majority of these irregulars occur at redshifts $1 < z < 2$.

~

\noindent
These conclusions are based on the powerful combination of morphological and 
photometric redshift data for all galaxies in the Hubble Deep Field to 
$I < 26$. Our interpretation is that the faint blue galaxy excess, which 
incidentally coincides with the peak in the 
observed mean galaxy star formation rate, represents the final formation epoch 
of the familiar spiral galaxy population. This conclusion is corroborated 
by the low abundance of normal spirals at $z > 2$. Taking these facts
together we favour a scenario where the faint blue excess is primarily due
to the formation epoch of spiral systems via merging at redshifts $1 < z < 2$.
The final interpretation now awaits refinements in our understanding of the 
{\it local} galaxy population.
\end{abstract}

\section{Introduction}
The faint blue galaxy problem has been with us for several decades and is 
comprehensively reviewed in Koo \& Kron (1992) and Ellis (1997). The problem
is surmised as: {\it an observed excess of faint galaxies over the zero
or passive evolution model predictions}. This excess first arises at
$b_{J} = 22$ mags and extends to the faintest magnitudes probed 
($b_{J} = 28.5$ mags, c.f. Metcalfe {\it et al.} 1995). The problem is 
compounded when one also considers the redshift distributions of galaxies at 
$b_{J}=22-24$ mags, e.g. Glazebrook {\it et al.} (1995a), which in
shape agree well with the zero-to-passive model predictions, but of course
not in amplitude ({\it i.e.} a reiteration of the original faint blue galaxy 
problem). Spectroscopic surveys to fainter magnitudes are 
currently limited by aperture and signal-to-noise considerations.
The unfortunate situation then, is that the models require a continuous 
renormalisation to match the observations and such a solution often results in 
contrived and implausible physical implications.
In this overview I summarise the recent substantial 
developments in the observational data from Hubble Space Telescope imaging and 
in particular the Hubble Deep Field (HDF), the current interpretation
and finally the future observations required.
The expectation is that through these refinements, as opposed 
to speculative retro-fitting, comes concordance.

\section{Hubble Deep Field Imaging}
The past three years have seen two large strides forward into new parameter 
space within this field, these are: the ability to discern morphologies 
(c.f. Odewahn {\it et al.} 1995) and the ability to reliably estimate 
redshifts/distances (c.f. Fern\'andez-Soto, these proceedings; 
Hogg {\it et al.} 1998).
Even more powerful however, is the combination of these two approaches and the 
generation of morphological number-count N(m,T), AND morphological 
redshift distributions N(z,T), to $I < 26$ (c.f. Driver {\it et al.} 1998). 
Figures 1 and 2 are adapted from Driver {\it et al.} and show the latest 
comparison between faint galaxy observations and models.
Before comparing the observations 
with the models it is first worth highlighting a number of purely observational
points:

\begin{enumerate}
\item At bright magnitudes the total counts are dominated by the 
classical Hubble types ({\it i.e.} ellipticals and spirals).
\item The elliptical galaxy counts are almost totally flat at faint
magnitudes and eventually contribute negligibly to the total galaxy counts.
\item The galaxy counts of disk and irregular systems are steep and almost 
linear over a broad magnitude range, with no sign of flattening and with 
gradients of $\sim 0.3$ and $\sim 0.4$ respectively.
\item At faint magnitudes the counts are divided almost equally between disks
and irregulars over a broad magnitude range.
\item All redshift distributions become broad towards fainter 
magnitudes and the Euclidean correlation between faintness and distance 
breaks down entirely.
\item Few spiral systems are seen at $z > 2$.
\item The late-type/irregular distributions are the broadest but typically
have a {\it higher} mean redshift then either of the so called ``giant'' 
classes.
\end{enumerate}

\begin{figure}[h]
\centerline{\psfig{file=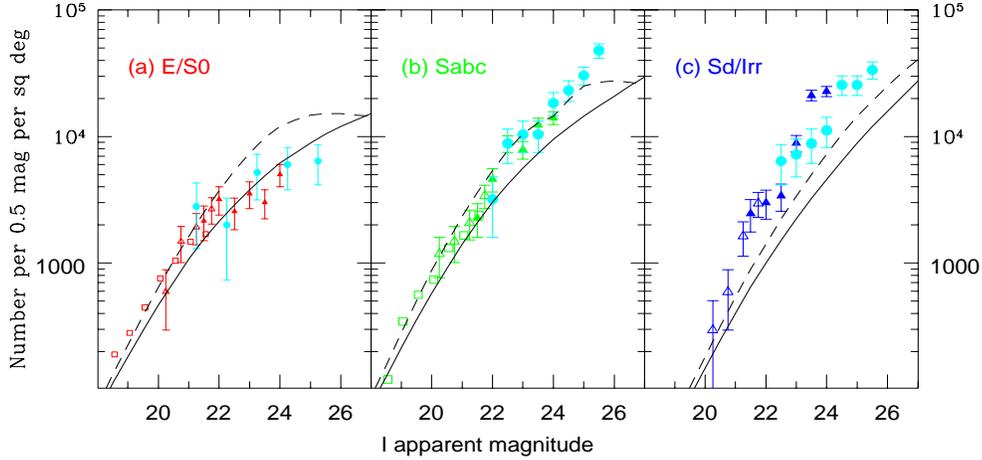,height=120mm,width=130mm}}
\vspace{-5.75cm}
\caption{Morphological galaxy number counts for: (a) all galaxies, (b)
ellipticals, (c) spirals, and (d) irregulars. The models lines show the
zero- (solid) and maximal passive- (broken) evolution models based on
a global renormalisation at $b_{J} = 18$.}
\end{figure}

\begin{figure}[p]
\centerline{\hspace{-6.0cm} \psfig{file=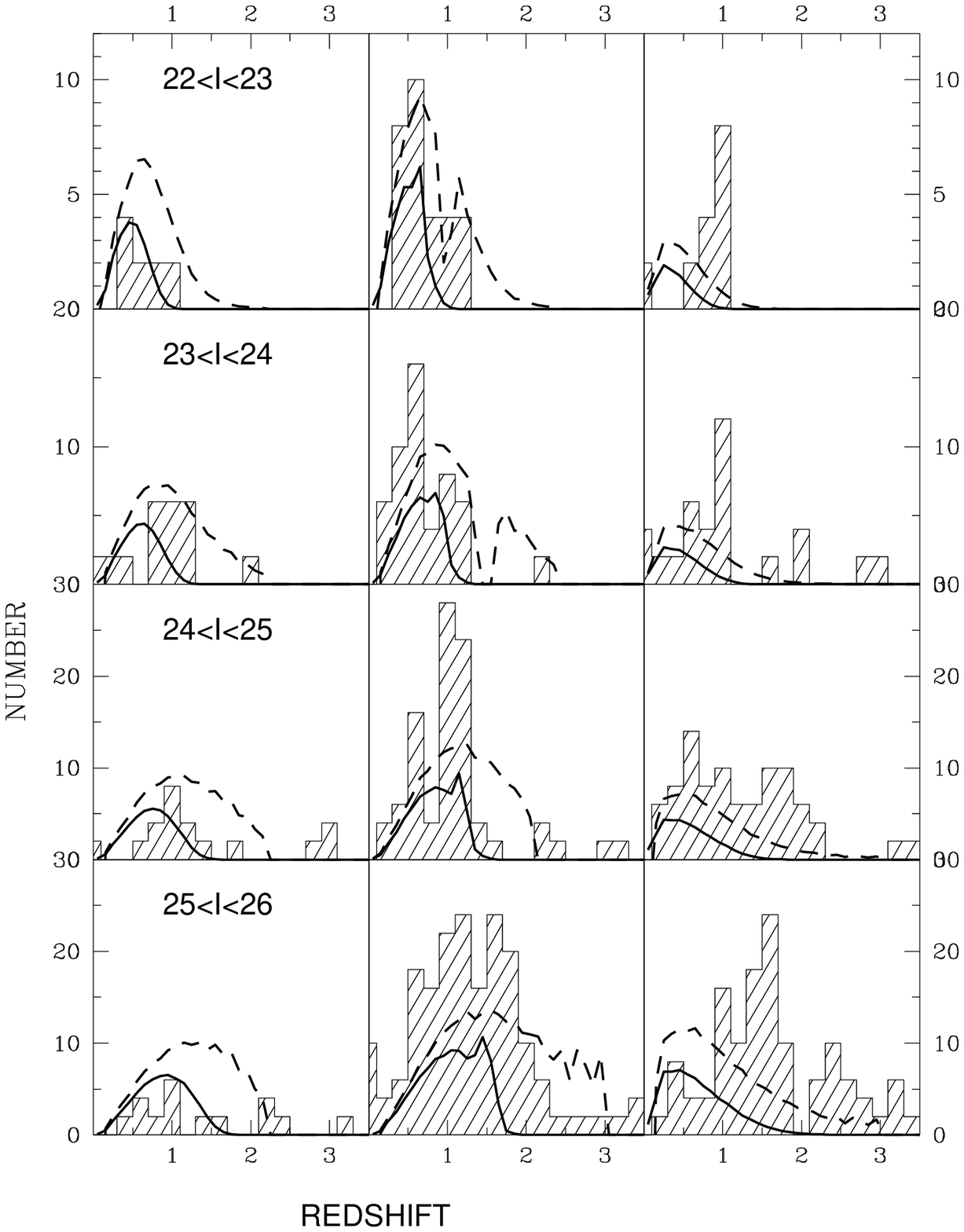,height=180mm,width=180mm}}
\caption{Redshift distributions for: (a) all galaxies, (b) ellipticals,
(c) spirals and (d) irregulars) for progressively fainter magnitude intervals.
The models lines show the zero- (solid) and maximal passive- (broken) evolution
models based on a global renormalisation at $b_{J}=18$.}
\end{figure}

At this point 
it is important to note that the data presented in these Figures is only 
feasible from a space borne imaging system, such as that 
onboard the Hubble Space Telescope.
Plate 1 shows a qualitative true colour representation of this same dataset 
subdivided into redshift intervals and arranged according to apparent magnitude
within these redshift intervals (and therefore crudely absolute magnitude).
Note that no correction can/has been made for the K-correction in this plate.
The plate essentially provides a good sanity check on the data quality and 
believability of the quantitative results presented in Figures 1 and 2. The 
trend toward higher irregularity at $z > 1.5$ is irrefutable and the degree of 
irregularity is far higher than that seen in the limited UV observations of 
local normal galaxies - {\it i.e.} the
first indication is that the increase in irregularity towards higher-z is 
intrinsic rather than a manifestation of the shifting bandpass.

\section{Faint Galaxy Modeling}
The models shown on Figures 1 and 2 are described in detail in Driver {\it et 
al.} (1998) and are based on the following: Local morphological luminosity 
functions from the CfA survey (c.f. Marzke {\it et al.} 1994); a standard
flat cosmology; zero or passive E- and K- corrections (c.f. Poggianti); 
and a global renormalisation at $b_{J} = 18$. By comparison with Figure 1 we 
see that the ellipticals are consistent with zero-evolution, spirals
with passive-evolution and irregulars require strong-evolution. These form the
basic conclusions derived in a number of previous papers based on 
morphological galaxy counts alone (e.g. Driver {\it et al.} 1995;
Glazebrook {\it et al.} 1995b) and independently from the Canada France
Redshift Survey (c.f. Lilly {\it et al.} 1995). This interpretation has
the obvious result that the level of evolution correlates with
the mean local colour of the respective galaxy sub-class, 
{\it i.e.} the systems with the oldest stellar populations 
require the least evolution and vica-versa. However with the inclusion of
photometric redshift data (described in Fern\'andez-Soto, these proceedings) we
can now take the comparison between observations and models one step further 
and, immediately, we see from Figure 2 that the description above is overly 
simplistic. Considering each class independently:

~

\noindent
The ellipticals' N(z,T=E/S0)s are well embraced by the zero-to-passive 
evolution N(z) models (bear in mind the strong clustering nature of 
ellipticals and the limiting statistics in this single sight-line). However, 
there is a tendency towards an underdensity of ellipticals in the faintest 
magnitude intervals suggesting some moderate obscurration, disassembly or 
transmorphing. Although taking into consideration the fact that the observed 
N(z,T=E/S0) distribution is
broad and not obviously truncated argues against a homogeneous evolutionary
path ({\it i.e.} no single epoch of formation !) and leads us to ask whether 
this population is formed via
a continuous ongoing crystallisation out of a non-elliptical population.
Given the limited statistics for this population more observations are 
required. A simpler interpretation is to simply adopt a lower normalisation 
for the elliptical models.

~

\noindent
For the spirals we see that passive-evolution models match both the counts and 
the brightest N(z) distribution. However, disparity creeps into the 
plots at fainter magnitude intervals. In fact there is a repetition
of the original faint blue galaxy problem within this sub-class in the 
sense that the observations agree in form to the zero-evolution models but are 
discrepant in amplitude. There are three
obvious possibilities: (1) spiral galaxies undergo zero- evolution and exhibit 
a steeper local luminosity function than that derived from the CfA; (2) 
spiral galaxies have disassembled into a number of similar luminosity
but less massive disk systems and (3) luminosity-dependen evolution. 
A frequent criticism leveled against
morphological classification at faint magnitudes is the concern that regular 
galaxies simply appear more asymmetrical due to the K-correction {\it i.e.}
objects viewed in the UV are intrinsically more irregular 
(e.g. Giavalisco {\it et al.} 1996). This is a valid
concern but there are two points worth noting: Firstly, from Plate 1 the 
irregulars at $z > 1.5$ are {\it extremely} irregular and secondly, even with 
the potential for mis-classification we still see too many spirals at lower 
redshifts implying substantial number evolution ($\times 2$).

~

\noindent
For the irregulars we saw from Figure 1, that even with the contentiously steep
CfA local luminosity function, extremely strong evolution would be required
before the model N(m,T=Sd/Irr) matches the observations.
However the N(z,T=Sd/Irr)s of Figure 2 make this a moot point, and it is clear 
that the irregular class is irreconcilable with the observed N(z,T=Sd/Irr) 
distribution. The implication is that more than
one population/process is contributing to the class of irregular galaxies: 
e.g. true irregulars --- typically low-luminosity; and transient irregulars 
--- peaking in the interval z = 1 - 2. Further work will be required to 
subdivide this population, however, before the nails are banged home in the 
dwarf-dominated coffin, we note that at no point does the N(z,T=Sd/Irr) model 
{\it strongly over predict} the observed distribution. 
Hence a steeply rising local luminosity function, 
akin to that seen in the CfA survey, is fully consistent and arguably favored 
by the low redshift end of the irregular N(z) distribution. 
However it is fair to say that the dwarf galaxy contribution to the faint 
galaxy counts is a minority component of the faint blue galaxy excess 
($ < 20\%$ to $I < 26$).

~

\noindent
Finally we must also consider a more holistic interpretation allowing freedom
of movement from one class to another. At some point all objects derive from
some primordial density distribution and the distinction between morphological 
classes must eventually disolve.
Considering the extreme distances over which the HDF objects in this dataset
are being seen this is clearly a serious consideration.
One interesting coincidence from Figure 2, is that the high-z 
irregulars are typically observed, within each magnitude interval, at a 
redshift slightly higher to the spiral population. This begs the question as 
to whether this population represents the spiral progenitor population 
(non-relaxed spirals) ! In fact one might argue to simply combine the spiral 
and irregular sub-classes in which case the overall form of the N(m,T=disk) 
distributions agree well with the models and we are simply left with a 
disk normalisation problem.

\section{What next ?}
A few things prevent a definitive description and these are
almost all due to uncertainties in our understanding of the local galaxy 
populations. The most important step forward over the next decade is to refocus
our attention on redefining local galaxy samples and this has been recognised
by the implementation of the 2dF and SLOAN surveys. However as has been shown
here the inclusion of morphologies is crucial and both of these
surveys need to find a way to incorporate such information. Accurate
consensus of the local morphological luminosity functions and their precise
normalisation would represent the single most important advancement. 
In fact it is the normalisation problem, not discussed here but see Shanks 
(1989), at $b_{J}=18$, which stymies our attempts at a definitive explanation.
The models shown here have been normalised uniformly to the total counts at
$b_{J} = 18.0$, which in the absence of morphological data is an
arbitrary decision taken to minimise the number of free parameters. If we
allow ourselves the luxury of independent morphological renormalisations then
we might adopt a renormalisation ratio of 1:2:3 for E/S0:Sabc:Sd/Irrs 
respectively. Such a normalisation goes a long way to simplifying the current
interpretation outlined in the previous section. If the reason for the local 
normalisation problem is
due to surface brightness selection effects as has been postulated then
such a renormalisation would be favoured. Once again we see that we require 
local morphological information at {\it brighter} rather than fainter 
magnitudes (where morphological distinctions are expected to evaporate).
How ironic it is then, that we now
know more about the statistical properties of the distant galaxy population
than that of the local population.

\noindent
We acknowledge and thank the engineers, scientists and astronauts 
responsible for the ongoing success of the {\it Hubble Space Telescope}.

\pagebreak
\begin{figure}
\vspace{20.0cm}
\caption{A colour montage of all galaxies in the Hubble Deep Field with 
$I < 26$, the galaxies are subdivided into redshift intervals as indicates
and according to apparent magnitude within each redshift interval. Note
the increase in irregularity towards both faint magnitudes (true 
irregulars) and high redshift (evolutionary irregulars). The increase
in irregularity at $z > 1.5$ 
coincides with the widely discussed ``Madau peak''.}
\end{figure}

\end{document}